%Paper: hep-ph/9402218
%From: Michael Earnshaw <MAE16@phx.cam.ac.uk>
%Date: Thu, 03 Feb 94 12:43:10 GMT

%Plain TeX
%
%
\overfullrule=0pt\magnification=\magstep1\baselineskip=20pt plus.1pt minus.1pt
\hsize=6.25 true in\vsize=9.65 true in\parindent=10 true pt
\font\reffont=cmcsc10\font\bigbf=cmbx10 scaled\magstep2
\def\newline{\par\noindent}
\def\startnumberingat#1{\pageno=#1\footline={\hss\tenrm\folio\hss}}
\newcount\eqnno\eqnno=0\newcount\secn\secn=0
\def\numeqn{\global\advance\eqnno by 1 \eqno(\the\eqnno)}
\def\rememeqn#1{\edef#1{\the\eqnno}}
\def\REF#1#2#3#4#5#6{\item{[#1]}{\reffont#2}{\it\ #3}{\bf\ #4}{\ (#5),\ #6}}

\newcount\refno\refno=0\def\ref#1{$\,^{[#1]}$}
\def\nextref#1{\global\advance\refno by 1\edef#1{\the\refno}}
\def\newsec#1\par{\vskip0pt plus .2\vsize\penalty-250
 \vskip0pt plus -.2\vsize\bigskip\vskip\parskip\global\advance\secn by 1
 \message {\the\secn. #1}\leftline{\bf\the\secn. #1}\nobreak\smallskip}
\def\beginsec#1\par{\vskip0pt plus .2\vsize\penalty-250\vskip0pt plus -.2\vsize
 \bigskip\vskip\parskip\message {#1}\leftline{\bf #1}\nobreak\smallskip}
\catcode`\@=11
\font\tenmsx=msxm10\font\sevenmsx=msxm7\font\fivemsx=msxm5\newfam\msxfam
\textfont\msxfam=\tenmsx\scriptfont\msxfam=\sevenmsx
\scriptscriptfont\msxfam=\fivemsx
\def\hexnumber@#1{\ifnum#1<10 \number#1\else
 \ifnum#1=10 A\else\ifnum#1=11 B\else\ifnum#1=12 C\else
 \ifnum#1=13 D\else\ifnum#1=14 E\else\ifnum#1=15 F\fi\fi\fi\fi\fi\fi\fi}
\def\msx@{\hexnumber@\msxfam}
\mathchardef\square="0\msx@03
\catcode`\@=12
\def\l{\vert l\vert}
\def\e{{\rm e}}
\pageno=0\nopagenumbers
\rightline{DAMTP 94--9}
\rightline{SWAT/23}
\rightline{hep-ph/9402218}
\bigskip\bigskip
\centerline{\bigbf STABILITY OF AN ELECTROWEAK STRING}
\centerline{\bigbf WITH A FERMION CONDENSATE}
\vskip 1 true in
\centerline{MICHAEL A. EARNSHAW$^1$}
\centerline{WARREN B. PERKINS$^2$}
\bigskip
\centerline{1.\ \it Department of Applied Mathematics and Theoretical Physics,}
\centerline{\it Silver Street, Cambridge, CB3 9EW, Great Britain}
\centerline{2.\ \it Department of Physics, University of Wales, Swansea}
\centerline{\it Singleton Park, Swansea, SA2 8PP}
\vskip 1 true in
\centerline{\hfill{\bf ABSTRACT}\hfill}\nobreak\smallskip\noindent

A solution of the standard electroweak theory with a single lepton
family is constructed, consisting of a cosmic string and a fermion
condensate within its core. The stability of this system to small
perturbations is examined, and it is found that stability is not
enhanced relative to the bare electroweak string. The presence of
quark zero modes is shown to violate the existence criteria for
embedded defects.

\vfill\eject
\startnumberingat1

\newsec INTRODUCTION

The electroweak phase transition has attracted a great deal of
\nextref\cohen
attention in recent years as a promising epoch for the generation of the
prim\ae val baryon asymmetry\ref\cohen. One mechanism examined in
detail in the
\nextref\turok
literature is that of baryon generation at the walls of bubbles of
false vacuum created during a first order phase transition\ref\turok.
\nextref\branden
Another proposal, which does not depend on the electroweak phase
transition being first order is that of the shrinking and eventual
collapse of strands of cosmic string\ref\branden. A common thread linking
both mechanisms is the requirement of a boundary between false and true
vacua moving in a definite direction.

\nextref\nielsen
It is well known that the Nielsen-Olesen vortex solution\ref\nielsen of the
abelian Higgs model may be embedded in the Weinberg-Salam electroweak
\nextref\vachA\nextref\nambu
theory; this new defect is a vortex of Z-particles, and is known as a
Z-string\ref{\vachA,\nambu}. To produce a sufficiently large baryon
asymmetry from
collapsing Z-strings, it is necessary to have a network of these
strings formed during the electroweak phase transition. The detailed
structure of such a network is difficult to analyse, but an important
factor in determining the length of string formed is the stability or
otherwise of such embedded strings. In contrast with vortices in the
abelian Higgs model, Z-strings are not topologically stable, and the
stability question becomes a dynamical one.

If one's analysis is restricted to the bosonic sector of the
\nextref\james\nextref\perkins
electroweak theory, the Z-string is found to be unstable in the physical
region of parameter space\ref{\james,\perkins}. Adding a second Higgs
doublet to
the model does not extend the region of parameter space admitting
\nextref\earnshaw
stable strings sufficiently to reach the physical values\ref\earnshaw.
On replacing
the leptonic sector of the theory, it is seen that leptonic zero modes
may be present along the string. A fermion whose mass arises from a
Yukawa coupling will be massless within the string core and massive
without. It has been suggested that, since the collapse of the string
will cause any fermion condensate within to become massive, the
existence of such a condensate will extend the region of parameter
\nextref\vachB
space admitting stable strings\ref\vachB. In the following work, this
conjecture is examined in a concrete model: the standard electroweak
theory with a single lepton family.

In section two, the structure of the Z-string and fermion condensate
is explicitly described. In section three this configuration
is perturbed, and the change in energy studied to second
order in the perturbations. The case of quark zero modes is discussed
in section four and the results are briefly summarised in
section five.

\newsec THE Z-STRING AND FERMION ZERO MODE

The model under investigation is the Weinberg-Salam electroweak theory
with a single lepton family, namely
$$\eqalign{{\cal L}=&-{1\over4}W^a_{\mu\nu}W^{a\mu\nu}-{1\over4}
F_{\mu\nu}F^{\mu\nu}+\left(D_\mu\Phi\right)^\dagger\left(D^\mu\Phi\right)
-\lambda\left(\Phi^\dagger\Phi-\eta^2\right)^2\cr
&-i{\bar\Psi}\gamma^\mu D_\mu\Psi-i{\bar e}_R\gamma^\mu D_\mu e_R
+h\left({\bar e}_R\Phi^\dagger\Psi+{\bar\Psi}\Phi e_R\right),\cr}
\numeqn$$
where
$$\eqalign{D_\mu\Psi&=\left(\partial_\mu-i{g\over2}\tau^a W^a_\mu
+i{{g'}\over2}B_\mu\right)\Psi,\cr
D_\mu e_R&=\left(\partial_\mu+ig'B_\mu\right)e_R,\cr
D_\mu\Phi&=\left(\partial_\mu-i{g\over2}\tau^a W^a_\mu
-i{{g'}\over2}B_\mu\right)\Phi,\cr}$$
$W^a_{\mu\nu}=\partial_\mu W^a_\nu-\partial_\nu W^a_\mu+g\epsilon^{abc}
W^b_\mu W^c_\nu$, $F_{\mu\nu}=\partial_\mu B_\nu-\partial_\nu B_\mu$, and
$\tau^a$ are the Pauli matrices. The field equations arising from this
Lagrangian are
$$\partial_\nu F^{\mu\nu}=i{{g'}\over2}
\left(\Phi^\dagger(D^\mu\Phi)-(D^\mu\Phi)^\dagger\Phi\right)+{{g'}\over2}
\left(\bar\Psi\gamma^\mu\Psi+2{\bar e}_R\gamma^\mu e_R\right),\numeqn$$
$$(\delta^{ab}\partial_\nu+g\epsilon^{abc}W_\nu^c)W^{a\mu\nu}
=i{g\over2}\left(\Phi^\dagger\tau^b(D^\mu\Phi)-(D^\mu\Phi)^\dagger
\tau^b\Phi\right)-{g\over2}\bar\Psi\gamma^\mu\tau^b\Psi,\numeqn$$
$$D^\mu D_\mu\Phi+2\lambda\left(\Phi^\dagger\Phi-\eta^2\right)\Phi=
h{\bar e}_R\Psi,\numeqn$$
$$i\gamma^\mu D_\mu\Psi=h\Phi e_R,\numeqn$$
$$i\gamma^\mu D_\mu e_R=h\Phi^\dagger\Psi.\numeqn$$

A Z-string ansatz of the form $W^1_\mu=W^2_\mu=0$,
$\Phi={0\choose \phi}$, $\Psi={0\choose e_L}$ satisfies the field
equations for $W^1_\mu$ and $W^2_\mu$ and the remaining equations become
$$-\square A_\mu+\partial_\mu(\partial.A)=q\sin2\Theta_W
{\bar e}\gamma_\mu e,\numeqn$$
$$\eqalign{-\square Z_\mu+\partial_\mu(\partial.Z)=&-iq\left(
\phi^\dagger(D_\mu\phi)-(D_\mu\phi)^\dagger\phi\right)\cr
&+q\cos2\Theta_W{\bar e}_L\gamma_\mu e_L-2q\sin^2\Theta_W
{\bar e}_R\gamma_\mu e_R,\cr}\numeqn$$
$$D^\mu D_\mu\phi+2\lambda\left(\phi^\dagger\phi-\eta^2\right)\phi=h
{\bar e}_R e_L,\numeqn$$
where $q={1\over2}\alpha$, $Z_\mu=\cos\Theta_W W^3_\mu-\sin\Theta_W B_\mu$,
$A_\mu=\sin\Theta_W W^3_\mu+\cos\Theta_W B_\mu$, $g'=\alpha\sin\Theta_W$,
$g=\alpha\cos\Theta_W$ and $D_\mu\phi=(\partial_\mu-iqZ_\mu)\phi$.
Working in cylindrical polar coordinates $(r,\theta,z)$ and writing
$A^\mu=0$,
$$qZ^\mu=\left(0,-{{v(r)}\over r}{\vec e}_\theta\right),
\quad \phi=\eta f(r)e^{i\theta},\numeqn$$
\rememeqn\profile
leads to the slightly modified vortex profile equations
$$-v''+{1\over r}v'+2q^2\eta^2f^2(v-1)=rq^2(\cos2\Theta_W{\bar e}_L
\gamma^\theta e_L-2\sin^2\Theta_W{\bar e}_R\gamma^\theta e_R),\numeqn$$
\rememeqn\eqnv
$$-f''-{1\over r}f'+{1\over{r^2}}f(1-v)^2+2\lambda\eta^2f(f^2-1)=
{g\over\eta}e^{-i\theta}{\bar e}_R e_L,\numeqn$$
\rememeqn\eqnf
where the Dirac matrices are
$$\gamma^r=\pmatrix{0&e^{-i\theta}&0&0\cr -e^{i\theta}&0&0&0\cr
0&0&0&-e^{-i\theta}\cr 0&0&e^{i\theta}&0\cr},\quad\gamma^\theta=\pmatrix{
0&-ie^{-i\theta}&0&0\cr -ie^{i\theta}&0&0&0\cr 0&0&0&ie^{-i\theta}\cr
0&0&ie^{i\theta}&0\cr},$$
$$\gamma^0=\pmatrix{\tau^3&0\cr 0&-\tau^3\cr},\quad\gamma^z=\pmatrix{
0&1\cr -1&0\cr},\quad\gamma^5=\pmatrix{0&1\cr 1&0\cr}.$$
The boundary conditions on $f$ and $v$ are
$$f(0)=v(0)=0,\quad f(\infty)=v(\infty)=1.\numeqn$$

The four-component Dirac spinor $e$ is split into the two parts $e_L$
and $e_R$. Writing $e_L^T=(a,b,-a,-b)$, $e_R^T=(c,d,c,d)$ leads to
Dirac equations which may be solved using the ansatz
$$\eqalign{a=&\psi_1(r)e^{ikz-i\omega t+in\theta},\cr
b=&i\psi_2(r)e^{ikz-i\omega t+i(n+1)\theta},\cr
c=&\psi_3(r)e^{ikz-i\omega t+i(n-1)\theta},\cr
d=&i\psi_4(r)e^{ikz-i\omega t+in\theta},\cr}\numeqn$$
\rememeqn\phases
where $\psi_1$, $\psi_2$, $\psi_3$, $\psi_4$ are real. This ansatz
results in the coupled ordinary differential equations
$$\eqalign{
(\omega+k)\psi_1-\psi_2'-{1\over r}(n+1+\cos2\Theta_W v)\psi_2&=h\eta
f\psi_3,\cr
(k-\omega)\psi_2-\psi_1'+{1\over r}(n+\cos2\Theta_W v)\psi_1&=h\eta
f\psi_4,\cr
(\omega-k)\psi_3-\psi_4'+{1\over r}(n-2\sin^2\Theta_W v)\psi_4&=h\eta
f\psi_1,\cr
-(\omega+k)\psi_4-\psi_3'+{1\over r}(n-1-2\sin^2\Theta_W v)\psi_3&=h\eta
f\psi_2.\cr}\numeqn$$
\rememeqn\diraca
We are interested in finding zero energy condensates, that is solutions to
equations (\diraca) with $\omega=k=0$ such that all the spinor
components decay exponentially at large $r$ and remain bounded in the
string core. On setting $\omega=k=0$, the equations decouple into two
pairs, namely
$$\eqalign{\psi_1'-{1\over r}(n+\cos2\Theta_W v)\psi_1&=-h\eta f\psi_4,\cr
\psi_4'+{1\over r}(n-2\sin^2\Theta_W v)\psi_4&=-h\eta f\psi_1,\cr}\numeqn$$
\rememeqn\diracb
and
$$\eqalign{\psi_2'+{1\over r}(n+1+\cos2\Theta_W v)\psi_2&=-h\eta f\psi_3,\cr
\psi_3'-{1\over r}(n-1-2\sin^2\Theta_W v)\psi_3&=-h\eta f\psi_2.\cr}\numeqn$$
\rememeqn\diracc
To match the exponentially decaying large $r$ solution in a given pair
\nextref\jackiw
with solution at the origin it is required that both solutions at
small $r$ are non-singular\ref\jackiw.
If the right hand sides of equations (\eqnv) and (\eqnf) are both
zero, then $f$ and $v$ are the familiar Nielsen-Olesen profiles, with
the limiting behaviour $f\propto r$ and $v\propto r^2$ for small $r$.
Assuming this limiting behaviour remains the same, it is found that,
whatever value of $n$ is chosen, the pair of equations (\diracc)
always has at least one singular solution for small $r$, hence we
must set $\psi_2=\psi_3=0$ to avoid exponentially large profiles at
large $r$ or singular behaviour at the origin. For the pair of
equations (\diracb), both solutions for small $r$ are non-singular iff
$n=0$, yielding a single leptonic zero energy condensate of the form
$$e_L=\pmatrix{1\cr0\cr-1\cr0\cr}\psi_1(r),\quad e_R=\pmatrix
{0\cr1\cr0\cr1\cr}i\psi_4(r),\numeqn$$
\rememeqn\zcond
where
$$\eqalign{\psi_1'-{1\over r}\cos2\Theta_W v\psi_1&=-h\eta f\psi_4,\cr
\psi_4'-{2\over r}\sin^2\Theta_W v\psi_4&=-h\eta f\psi_1.\cr}\numeqn$$
With this form of the condensate, equations (\eqnv) and (\eqnf) reduce to
$$-v''+{1\over r}v'+2q^2\eta^2f^2(v-1)=0,\numeqn$$
$$-f''-{1\over r}f'+{1\over{r^2}}f(1-v)^2+2\lambda\eta^2f(f^2-1)=0,
\numeqn$$
thus the Higgs field and gauge field profiles are identical to those
for the usual Z-string, and the earlier assumptions about their limiting
behaviour for small $r$ are valid.

Multiplying the spinors given in equation (\zcond) by $e^{-i\omega(t+z)}$
for any value of $\omega$ gives a zero mode moving along the string
which is still a solution of equations (\diraca).

It should be noted that all the field equations are satisfied, except
those for $A_0$, $A_z$, $Z_0$ and $Z_z$, so the simple ansatz above
only yields an approximate solution. A reduction of the problem to two
space dimensions would remove two of these inconsistencies, still
leaving problems with $A_0$ and $Z_0$, but would spoil the
decomposition of the electron field into left and right handed
components, which relies on there being exactly three spatial dimensions.

\newsec STABILITY OF THE Z-STRING WITH A LEPTON CONDENSATE

Since the vacuum manifold of the bosonic electroweak theory is $S^3$,
cosmic string configurations are not topologically stable, but if the
string solution is meta-stable, that is a local minimum of energy in
configuration space, then such strings may have significant
cosmological consequences. It has been shown that the Z-string is
meta-stable in a region of parameter space near $\Theta_W=\pi/2$, but
not at physical values of the parameters\ref\james. In this section, the
configuration described in section two is peturbed infinitesimally,
and the quadratic change in the energy of the system is examined to
discover whether or not the fermion condensate enhances the stability
properties.

The energy of a {\it static} configuration in the model under investigation
is
$$\eqalign{E=\int d^3x\Big\{&{1\over4}W^a_{ij}W^a_{ij}+{1\over4}F_{ij}F_{ij}
+(D^i\Phi)^\dagger(D^i\Phi)+\lambda\left(\Phi^\dagger\Phi-\eta^2\right)^2\cr
&-i\bar\Psi\gamma^iD^i\Psi-i{\bar e}_R\gamma^iD^ie_R-h({\bar e}_R
\Phi^\dagger\Psi+\bar\Psi\Phi e_R)\Big\}.\cr}\numeqn$$
\rememeqn\energy
For simplicity, perturbations will be restricted to be independent of
$z$, and the $z$-components of the gauge field perturbations will be
set to zero, thus allowing a study of energy per unit length. The main
conclusion concerning stability of the system is unaffected by
this restriction, as will be explained later. Writing
$$\Phi=\pmatrix{\chi\cr\phi+\delta\phi\cr},\quad\Psi=\pmatrix{\delta\nu_L
\cr e_L+\delta e_L\cr},\numeqn$$
and replacing $\vec Z$ by $\vec Z+\delta\vec Z$, $\vec W^{1,2}$ by $\delta
\vec W^{1,2}$, $\vec A$ by $\delta\vec A$, and $e_R$ by $e_R+\delta
e_R$, where $\vec Z$, $\phi$, $e_L$, and $e_R$ are given by expressions
(\profile) and (\zcond), the change in energy per unit length to
quadratic order in the perturbations is found to be of the form
$$\delta E=\delta E_1(\delta\vec W^1,\delta\vec W^2,\chi,\delta\nu_L)
+\delta E_2(\delta\phi,\delta\vec Z,\delta\vec A,\delta e_L,\delta e_R)
\numeqn$$
\rememeqn\energysplit
where
$$\delta E_1=\delta E_1^B(\delta\vec W^1,\delta\vec W^2,\chi)
+\delta E_1^F(\delta\vec W^1,\delta\vec W^2,\chi,\delta\nu_L),\numeqn$$
$$\delta E_2=\delta E_2^B(\delta\phi,\delta\vec Z,\delta\vec A)
+\delta E_2^F(\delta\phi,\delta\vec Z,\delta\vec A,\delta e_L,\delta e_R).
\numeqn$$
Detailed expressions for $\delta E_1^B$, $\delta E_1^F$, $\delta
E_2^B$ and $\delta E_2^F$ are
$$\eqalign{\delta E_1^B=\int d^2x\Big\{&{1\over2}\vert\nabla\times\delta
\vec W^1+g\cos\Theta_W\vec Z\times\delta\vec W^2\vert^2+{1\over2}\vert\nabla
\times\delta\vec W^2-g\cos\Theta_W\vec Z\times\delta\vec W^1\vert^2\cr
&-g\cos\Theta_W\delta\vec W^1\times\delta\vec W^2.\nabla\times\vec Z
+{1\over4}g^2\phi^\dagger\phi\delta\vec W^\dagger.\delta\vec W\cr
&+\vert\nabla\chi\vert^2+2\lambda(\phi^\dagger\phi-\eta^2)\chi^\dagger\chi
+{1\over4}\alpha^2\cos^22\Theta_W\vec Z^2\chi^\dagger\chi\cr
&-{1\over4}g\alpha\sin^2\Theta_W\vec Z.(\delta\vec W\phi\chi^\dagger
+\delta\vec W^\dagger\phi^\dagger\chi)+{i\over2}\alpha\cos2\Theta_W
\vec Z.(\chi\nabla\chi^\dagger-\chi^\dagger\nabla\chi)\cr
&+{i\over2}g\left(\delta\vec W.(\phi\nabla\chi^\dagger-\chi^\dagger\nabla\phi)
+\delta\vec W^\dagger.(\chi\nabla\phi^\dagger-\phi^\dagger\nabla\chi)\right)
\Big\},\cr}\numeqn$$
$$\eqalign{\delta E_1^F=\int d^2x\Big\{&
i\delta\bar\nu_L\vec\gamma.\nabla\delta\nu_L+{1\over2}\alpha\delta\bar\nu_L
\vec\gamma.\vec Z\delta\nu_L\cr
&-{1\over2}g(\delta\bar\nu_L\vec\gamma.\delta\vec We_L+\bar e_L\vec\gamma.
\delta\vec W^\dagger\delta\nu_L)-h(\delta\bar\nu_L\chi e_R+\bar e_R
\chi^\dagger\delta\nu_L)\Big\}\cr}\numeqn$$
and
$$\eqalign{\delta E_2^B=\int d^2x\Big\{&
{1\over2}\vert\nabla\times\delta\vec Z\vert^2+{1\over4}\alpha^2
\phi^\dagger\phi\delta\vec Z^2
+{1\over2}\vert\nabla\times\delta\vec A\vert^2\cr
&+\vert\nabla\delta\phi\vert^2+2\lambda(\phi^\dagger\phi-\eta^2)\delta
\phi^\dagger\delta\phi
+\lambda(\phi^\dagger\delta\phi+\delta\phi^\dagger\phi)^2\cr
&+{1\over4}\alpha^2(\vec Z^2\vert\delta\phi\vert^2+2\vec Z.\delta\vec Z
(\phi^\dagger\delta\phi+\phi\delta\phi^\dagger))\cr
&+{i\over2}\alpha\delta\vec Z.(\delta\phi^\dagger\nabla\phi-\delta\phi
\nabla\phi^\dagger+\phi^\dagger\nabla\delta\phi-\phi\nabla\delta
\phi^\dagger)\cr
&+{i\over2}\alpha\vec Z.(\delta\phi^\dagger\nabla\delta\phi-\delta\phi
\nabla\delta\phi^\dagger)
\Big\},\cr}\numeqn$$
$$\eqalign{\delta E_2^F=\int d^2x\Big\{&
i\delta\bar e_L\vec\gamma.\nabla\delta e_L+i\delta\bar e_R\vec\gamma.\nabla
\delta e_R+{1\over2}\alpha\cos2\Theta_W\delta\bar e_L\vec\gamma.\vec
Z\delta e_L-\alpha\sin^2\Theta_W\delta\bar  e_R\vec\gamma.\vec Z\delta e_R
\cr&+{1\over2}\alpha(\cos2\Theta_W\delta\vec Z+\sin2\Theta_W\delta\vec A).
(\bar e_L\vec\gamma\delta e_L+\delta\bar e_L\vec\gamma e_L)\cr
&+g'(\cos\Theta_W\delta\vec A-\sin\Theta_W\delta\vec Z).(\bar e_R\vec\gamma
\delta e_R+\delta\bar e_R\vec\gamma e_R)\cr
&-h(\phi^\dagger\delta\bar e_R\delta e_L+\phi\delta\bar e_L\delta e_R
+\delta\phi^\dagger\bar e_R\delta e_L+\delta\phi\delta\bar e_Le_R
+\delta\phi^\dagger\delta\bar e_Re_L+\delta\phi\bar e_L\delta e_R)
\Big\},\cr}\numeqn$$
where $\delta\vec W=\delta\vec W^1-i\delta\vec W^2$.

It should be noted that a similar perturbative expansion in the pure
bosonic theory yields the same functional form as $\delta E_1^B+\delta E_2^B$.
Since the $\vec Z$ and $\phi$ profiles for the Z-string are unchanged
on addition of the fermion condensate constructed in section two, it
thus follows that
$$\delta E^B=\delta E_1^B+\delta E_2^B.\numeqn$$
The pure bosonic theory has been studied in detail by James et al.\ref\james,
who determined the region of $(\lambda,\Theta_W)$ space in which the
strings were unstable by explicitly constructing perturbations
yielding a negative energy change to quadratic order. With the fermion
condensate present, setting
$$\delta\nu_L=\delta e_L=\delta e_R=0\numeqn$$
\rememeqn\nofermpert
causes $\delta E_1^F$ and $\delta E_2^F$ to become zero, hence
$$\delta E=\delta E^B$$
for perturbations of this special type.

Suppose the parameters $\lambda$ and $\Theta_W$  are chosen such that
the Z-string in the pure bosonic theory is unstable, and the point in
parameter space is not on the boundary between stability and
instability. Then one can construct a perturbation of the form
$$\chi=\chi(x,y),\quad\delta\vec W=(W_x(x,y),W_y(x,y),0),\quad\delta\phi=0
,\quad\delta\vec Z=\delta\vec A=0$$
with negative $\delta E^B$. Using this perturbation, along with the zero
fermionic perturbations given in equation (\nofermpert), $\delta E$ is
found to be negative too, hence the region of $(\lambda,\Theta_W)$
parameter space yielding stable Z-strings with a fermion condensate is
contained in that yielding stable Z-strings.

The full problem with $z$ dependence contains the sub-problem of
energy change per unit length discussed above as a special case,
and the negative energy perturbation constructed above can be embedded
in the full perturbation expression yielding the same change
in energy per unit length. Thus, the region of stability in the
complete problem is contained within that of the Z-string with no condensate.

\newsec QUARK ZERO MODES

In the previous sections we have seen that the appearance of electron
zero modes on the electroweak string does not enhance the stability of
the vortex solution.  The lack of backreaction on the vortex fields is
due to the special form of the zero mode spinors (\zcond).  The nature of
the spinors is determined by the form of the coupling of the fermion to
the scalar field. This question is now considered in relation to the
appearance of quark zero modes.

Consider a generalisation of the string background fields (\profile) to
allow the string to have winding number $l$; the scalar field then has
an angular dependence of the form $\e^{il\theta}$. To remove the
angular variation from the Dirac equations for $\Psi$ and $e_R$ we must
modify our ansatz for the spinor components (\phases) by including an
extra factor of $\e^{-il\theta}$
in the definitions of $c$ and $d$ (the ansatz (\phases) is given for the
case $l=1$). The equations still decouple into two pairs on setting
$w=k=0\,$; equations (\diracb) and (\diracc) become
$$\eqalign{\psi_1'-{1\over r}(n+\cos2\Theta_W v)\psi_1&=-h\eta f\psi_4,\cr
\psi_4'+{1\over r}(n+1-l-2\sin^2\Theta_W v)\psi_4&=-h\eta f\psi_1,\cr}
\numeqn$$
\rememeqn\diracbq
and
$$\eqalign{\psi_2'+{1\over r}(n+1+\cos2\Theta_W v)\psi_2&=-h\eta f\psi_3,\cr
\psi_3'-{1\over r}(n-l-2\sin^2\Theta_W v)\psi_3&=-h\eta f\psi_2.\cr}\numeqn$$
\rememeqn\diraccq
At large distances we have the usual exponentially growing and
exponentially decaying solutions for massive fields. If we are to
have zero modes the small distance solutions must all be regular
so that we can construct a square integrable solution. Thus we need
only look at the case of small $r$. Close to the string core, $h\eta f
=sr^{\vert l\vert}$ for some constant $s$
 and $Z_\theta\propto r$, thus we can drop the gauge
field  terms and eliminate either spinor component of either pair
 to obtain a second order equation
for the  other.

Eliminating $\psi_4$ from (\diracbq), we find
$$
-{\l\over r}(-\psi'_1 +{n\over r}\psi_1) -\psi''_1 -{1-l\over r}\psi'_1
+{n\over r}{n-l\over r}\psi_1=0.\numeqn$$
Substituting $\psi_1=r^t$, yields the condition
$$-\l(n-t)-t^2+lt+n(n-l)=0.$$
If $\l=-l$ the
solutions are $t=\pm n$ and we have one regular and one irregular solution.
However, if $\l=l$ the solutions are $t=n$ and $t=-n+2l$ which can both be
regular if $l$ is nonzero.
Similarly we can eliminate $\psi_1$ and on
substitutng $\psi_4=r^t$, we obtain
$$\l(t+n+1-l)-t^2+lt+(n-l)(n+1-l)=0.$$
If $\l=-l$ the solutions are $t=\pm (n+1-l)$ and we again have one regular
and one irregular solution.
However, if $\l=l$ the solutions are $t=l\pm(n+1)$ which can also both be
regular if $l$ is nonzero.

Repeating this procedure for the other pair of fields we find that zero
modes with nonvanishing $\psi_2$ and $\psi_3$ only occur for $l$ negative.
Thus if $l$ is positive we have a zero mode of the form given in (\zcond)
and for $l$ negative we have modes of the form
$$
e_L=\pmatrix{0\cr1\cr 0\cr -1\cr}\psi_2(r),\quad e_R=\pmatrix
{1\cr0\cr1\cr0\cr}i\psi_3(r),\numeqn$$
\rememeqn\zcondq
For $l$ positive the modes persist if we allow $w$ and $k$ to be nonzero
but set $w+k=0$. The corresponding condition for $l$ negative is $w=-k$.
The sign of $l$ thus determines the direction of motion of the zero modes
along the string and the form of the spinor.
This will be important when studying quark zero modes.

Now consider introducing the up and down quarks to complete the first family
of fermions. We need to introduce mass terms for both right handed quarks
and so have a quark Lagrangian of the form
$$\eqalign{
{\cal L}_{\rm quark}= &
(\bar u,\bar d)_L \gamma^\mu(-i\partial_\mu-{g\over2}\tau^a W_\mu^a
+{g'\over 6} B_\mu) \pmatrix{u\cr d\cr}_L
+\bar u_R\gamma^\mu(-i\partial_\mu-{2\over 3}g'B_\mu)u_R
\cr &
+\bar d_R\gamma^\mu(-i\partial_\mu+{1\over 3}g'B_\mu)d_R
-G_d\biggl[
(\bar u,\bar d)_L \pmatrix{\phi^+\cr \phi\cr} d_R +\bar d_R(\phi^-,
\bar\phi) \pmatrix{u\cr d\cr}_L \biggr]
\cr&
-G_u\biggl[
(\bar u,\bar d)_L \pmatrix{-\bar\phi\cr \phi^-\cr} u_R +\bar u_R(-\phi,
\phi^+) \pmatrix{u\cr d\cr}_L \biggr],
\cr}\numeqn$$
\rememeqn\qlag
where $\Phi={\phi^+\choose\phi}$.
In the background fields of the electroweak string this reduces to
$$\eqalign{
{\cal L}_{\rm quark}= &
\bar u_L \gamma^\mu(-i\partial_\mu -{1 \over 2G}(g^2+g'^2/3)Z_\mu)u_L +
\bar d_L \gamma^\mu(-i\partial_\mu +{1 \over 2G}(g^2+g'^2/3)Z_\mu)d_L
\cr &
+\bar u_R\gamma^\mu(-i\partial_\mu+{2\over 3}{g'^2\over G} Z_\mu)u_R
+\bar d_R\gamma^\mu(-i\partial_\mu-{1\over 3}{g'^2\over G} Z_\mu)d_R
-G_d(\bar d_L\phi d_R +\bar d_R \phi^* d_L)
\cr &
+G_u(\bar u_L\phi^* u_R + \bar u_R\phi u_L).
\cr}\numeqn$$
The generic form of the Lagrangian for each pair of particles is thus
$$
{\cal L}_{\rm f}=
\bar f_L \gamma^\mu(-i\partial_\mu -aZ_\mu)f_L
+\bar f_R\gamma^\mu(-i\partial_\mu -bZ_\mu)f_R
+m(\bar f_L\phi^* f_R + \bar f_R\phi f_L),
\numeqn$$
where $\phi$ should be replaced by $\phi^*$ in the case of the d quark and
electron.

\nextref\witten
The coupling of the up quark to $\phi^*$ rather than $\phi$ leads to
the up quark zero modes moving in the opposite direction along the string
to the electron and down quark modes\ref\witten with
the form of the up quark
mode being given by (\zcondq) rather than (\zcond). Now,
the quark Lagrangian (\qlag) contains terms that are linear in the upper
component of the Higgs field: $-\phi^+(G_d\bar u_L d_R+G_u\bar u_R d_L) +
({\rm h.c.})$.
As the u and d quarks couple to the Higgs field in different
ways, one is a left-mover and the other a right-mover. The terms that can
act as
sources for the upper component of the Higgs doublet thus contain both
left and right handed and left and right moving fermions. In the gamma
basis
we are using, $\gamma^0=$ diag(1,-1,-1,1), thus using $\uparrow$ and
$\downarrow$ to denote left and right movers, the source terms are
$$
\bar f_L^\uparrow f_R^\downarrow=(0,\psi_2^*,0,-\psi_2^*)\gamma^0
\pmatrix {0\cr \psi_4\cr 0\cr \psi_4\cr}=2\psi_2^* \psi_4\numeqn
$$
and
$$
\bar f_R^\uparrow f_L^\downarrow=(\psi_3^*,0,\psi_3^*,0)\gamma^0
\pmatrix {\psi_1\cr  0\cr -\psi_1\cr 0 \cr}=2\psi_3^* \psi_1.
\numeqn$$
Thus the combination of u and d quarks will act as a source for the
upper component of the Higgs field, violating the Vachaspati existence
criteria\ref\vachA. It is only possible to write down the electroweak
string as a
solution of the equations of motion if there are no terms in the Lagrangain
that are linear in any of the fields that are assumed to vanish. The
quark zero modes violate this condition by generating a non-vanishing
term that is linear in the upper component
of the Higgs doublet.

\newsec SUMMARY

A zero energy leptonic bound state has been constructed, and it is
found that the profile of the Z-string is unaffected by its presence,
in the approximation of neglecting source terms in the $z$-direction and
the zero-direction. Negative energy perturbations for the bare
Z-string remain so for the configuration of Z-string and condensate,
and so the region of stability in parameter space is not extended -- in
particular, the configuration is unstable for physical values of the
electroweak parameters.

The presence of a fermionic zero mode implies that the electroweak
Z-string is superconducting. For this to have cosmological
significance, it is necessary to form such a string and then allow the
fermionic current to build up. As the string is unstable, even in the
presence of a time-independent fermion condensate, such a
configuration will not have been able to form at the electroweak phase
transition, and so the attractive scenario of a network of stabilised
loops of current-carrying electroweak string persisting to the present
day is unrealistic in this model.

The appearance of quark zero modes on the electroweak string leads to
a violation of the Vachaspati existence criteria for embedded defects
as the quark modes provide a source for the upper component of the
Higgs doublet.

\beginsec ACKNOWLEDGEMENTS

We would like to thank A-C. Davis for useful comments. This work was
funded by a S.E.R.C. research grant (M.A.E.) and a University of Wales
collaborative research grant(W.B.P.).

\beginsec REFERENCES

\frenchspacing

\item{[\cohen]}{\reffont A. Cohen, D. Kaplan, A. Nelson}{\ preprint
UCSD--PTH--93--02, hep-ph/9302210}

\REF{\turok}{N. Turok, J. Zadrozny}{Phys. Rev. Lett.}{65}{1990}{2331}

\REF{\branden}{R. Brandenberger, A-C. Davis}{Phys. Lett.}{B308}{1993}{79}

\REF{\nielsen}{H.B. Nielsen, P. Olesen}{Nucl. Phys.}{B61}{1973}{45}

\REF{\vachA}{T. Vachaspati}{Phys. Rev. Lett.}{68}{1992}{1263}

\REF{\nambu}{Y. Nambu}{Nucl. Phys.}{B130}{1977}{505}

\REF{\james}{M. James, T. Vachaspati, L. Perivolaropoulos}{Nucl. Phys}
{B395}{1993}{534}

\REF{\perkins}{W.B. Perkins}{Phys. Rev.}{D47}{1993}{5224}

\REF{\earnshaw}{M.A. Earnshaw, M. James}{Phys. Rev.}{D48}{1993}{5818}

\item{[\vachB]}{\reffont T. Vachaspati, R. Watkins}{\ preprint hep-ph/921184}

\REF{\jackiw}{R. Jackiw, P. Rossi}{Nucl. Phys.}{B190}{1981}{681}

\REF{\witten}{E. Witten}{Nucl. Phys.}{B249}{1985}{557}
\bye